\documentclass[amsfonts,amsmath]{revtex4}
\usepackage{graphicx}
\begin{document}
\def\BJ{{\mathcal{J}}}
\def\u{\boldsymbol}
\title{Thresholds to Chaos and Ionization for the Hydrogen Atom in Rotating Fields}
\date{\today}
\author{C. Chandre$^{1}$, David Farrelly$^{2}$, and T. Uzer$^{1}$}
\affiliation{$^{1}$Center for Nonlinear Science, School of Physics,
Georgia Institute of Technology, Atlanta, Georgia 30332-0430\\
$^{2}$Department of Chemistry and Biochemistry, Utah State University,
Logan, Utah 84322-0300}

\begin{abstract}
We analyze the classical phase space of the hydrogen atom in crossed magnetic 
and circularly polarized microwave fields in the high frequency regime, u
sing 
the Chirikov resonance overlap criterion and the renormalization map. These 
methods are used to compute thresholds to large scale chaos and to ionization. 
The effect of the magnetic field is a strong stabilization of a set 
of invariant tori which bound the trajectories and prevent stochastic 
ionization. In order to ionize, larger amplitudes of the microwave field 
are necessary in the presence of a magnetic field. 
\end{abstract}     \maketitle

\section{Introduction}

The chaotic ionization of the hydrogen atom in a variety of external fields
is a fundamental problem in nonlinear dynamics and atomic physics. Early
work, in particular, focused on the ionization mechanism in a linearly
polarized (LP) microwave field~\cite{casa87,koch95}. This problem is noteworthy 
because it showed the general applicability of the ``Chirikov Resonance Overlap 
Criterion''~\cite{chir79} to real quantum mechanical systems. This empirical 
criterion predicts the value at which two nearby resonances overlap and large 
scale stochasticity occurs. There is no doubt that the Chirikov criterion is 
generally robust and, therefore, it is no surprise that Chirikov's criteria has 
been tried as a way to predict transitions to chaos and ionization dynamics in 
more complicated circumstances, e.g., for the hydrogen atom in circularly 
polarized (CP)\ microwave fields. However, the success of the Chirikov criterion 
in describing the LP problem stands in contrast with what seems to have been a 
somewhat mixed performance when applied to the hydrogen atom in rotating 
microwave fields. Partly for this reason, a good deal of 
controversy has surrounded the ionization mechanism and also the 
interpretation of the Chirikov criterion when applied to ionization in 
rotating fields. In this article we propose to examine the ionization of the 
hydrogen atom in a circularly polarized microwave field (CP). 
Our analysis will use an advanced method, the renormalization 
map~\cite{koch99,chan01R}, so as to include higher order resonances beyond the 
Chirikov approach.

Interest in the CP\ microwave problem began with experiments by the Gallagher
group~\cite{fu90} which revealed a strong dependence of the ionization
threshold on the degree of polarization. They
explained their CP results proposing that in a rotating frame, ionization 
proceeds in roughly the same way as for a static field, i.e., a static field has 
the same effect whether its coordinate system is rotating or not. In a Comment, 
Nauenberg~\cite{naue90} argued that the ionization mechanism was substantially 
more complicated and that the effect of rotation on the ionization threshold had 
to be taken into account. Fully three-dimensional numerical simulations by 
Griffiths and Farrelly~\cite {grif92} were able to provide quantitative 
agreement with the experimental results. Griffiths and Farrelly~\cite{grif92} 
and Wintgen~\cite{wint91} independently proposed similar models for ionization 
based on the Runge-Lenz vector. \\

Subsequent theoretical work can be divided into three main categories: 
$(i)$ classical simulations~\cite{kapp93}, $(ii)$ quantum simulations, and 
$(iii)$ resonance overlap studies. The two main papers on resonance overlap are 
by Howard~\cite{howa92} who published the first such study of this system, and 
by Sacha and Zakrzewski~\cite {sach97}. In both cases, the Hamiltonian was 
written in an appropriate rotating frame, a choice of `zero-order' actions was 
made and the Chirikov machinery invoked. Somewhat surprisingly, the paper by 
Sacha and Zakrzewski~\cite{sach97} disagrees in a number of key areas with the 
results of Howard~\cite{howa92} as well as with some conclusions drawn from 
numerical studies \cite{farr95}. In a paper by Brunello {\it et 
al.}~\cite{brun97}, it was shown both numerically and analytically that the 
actual ionization threshold observed in an experiment must take into account the 
manner in which the field was turned on; essentially, in some experiments 
electrons are switched during the field ``turn on'' directly into unbound 
regions of phase space. That is, the underlying resonance structure of the 
Hamiltonian is almost irrelevant because ionization is accomplished by the 
ramp-up of the field. For this reason, the application of resonance overlap 
criteria in rotating frames can be quite intricate and this provides some 
explanation for the apparent inadequacy of the Chirikov criterion - if 
ionization occurs during the field turn-on time then, of course, resonance 
overlap is irrelevant.

In this article, we study the hydrogen atom driven by a CP
microwave field together with a magnetic field lying perpendicular to
the polarization plane (CP$\times B$). The magnetic field has been introduced to 
prevent ionization in the plane. This provides the opportunity to eliminate 
difficulties associated with the turn on of the field, thus opening up the way 
to a study of resonance overlap in a more controlled manner in the rotating 
frame. As noted, without the added magnetic field all the electrons may have 
gone before the resonances have had a chance to overlap.

The paper is organized as follows: Section II introduces the Hamiltonian and
provides a discussion of resonance overlap in the rotating frame in both the
Chirikov and renormalization approaches. In order to obtain qualitative and
quantitative results about the dynamics of the hydrogen atom in CP$\times $B
fields, we compare numerically, in Sec.\ III, Chirikov's resonance overlap 
criterion with our renormalization group transformation method. The Chirikov 
method is found to provide good qualitative results and is useful because of its 
simplicity. The renormalization transformation is used first to check the 
qualitative features obtained by the resonance overlap criterion, and to obtain 
more quantitative results about the dynamics by expanding the Chirikov approach. 
\ Conclusions are in Sec. IV.

\section{Hydrogen atom in CP$\times$B fields}

\label{sec:mode}

We consider a hydrogen atom perturbed by a microwave field of amplitude $F$
and frequency $\omega_f$, circularly polarized in the orbital plane, and a
magnetic field with cyclotron frequency $\omega_c$. The Hamiltonian in
atomic units and in the rotating frame of the microwave field is~\cite{Bfrie91}~: 
\begin{equation}
H(p_x,p_y,x,y)=\frac{p_x^2+p_y^2}{2}- \frac{1}{\sqrt{x^2+y^2}}-\Omega(xp_y-y
p_x)+F x +\frac{\omega_c^2}{8}(x^2+y^2),  \label{eqn:tro}
\end{equation}
where $\Omega=\omega_f-\omega_c/2$. Following Ref.~\cite{howa92}, we rewrite
Hamiltonian~(\ref{eqn:tro}) in the action-angle variables $(J,L,\theta,\psi)$
of the problem with $\omega_c=0$. The angles $\theta$ and $\psi$ are
conjugate respectively to the principal action $J$ and to the angular
momentum $L$. The Hamiltonian becomes~: 
\begin{equation}  \label{eqn:ham}
H(J,L,\theta,\psi)=H_0(J,L,\theta)+ FJ^2\sum_{n=-\infty}^{+\infty}
V_n(e)\cos (n\theta+\psi),
\end{equation}
where the integrable Hamiltonian $H_0$ is 
\begin{equation}
H_0(J,L,\theta)=-\frac{1}{2J^2}-\Omega L +J^4 \frac{\omega_c^2}{8}%
\sum_{n,m=-\infty}^{+\infty} V_m V_{m-n} \cos n\theta .
\end{equation}
The coefficients $V_n$ of the expansion are the following ones~: 
\begin{eqnarray}
&& V_0(e)=-\frac{3e}{2}, \\
&& V_n(e)=\frac{1}{n}\left[ {\cal J}^{\prime}_n(ne)+ \frac{\sqrt{1-e^2}}{e}
{\cal J}_n(ne)\right], \, \mbox{ for } n\not= 0,
\end{eqnarray}
where ${\cal J}_n$ is the $n$th Bessel function of the first kind and ${\cal 
J}_n^{\prime}$ its derivative. The parameter $e$ is given by 
\[
e=\left(1-\frac{L^2}{J^2}\right)^{1/2}. 
\]
The positions $x$ and $y$ are given by the following formulas~: 
\begin{eqnarray*}
&& x=\sum_{n=-\infty}^{+\infty} V_n(e) \cos (n\theta+\psi), \\
&& y=\sum_{n=-\infty}^{+\infty} V_n(e) \sin (n\theta+\psi).
\end{eqnarray*}
The Hamiltonian~(\ref{eqn:ham}) can be rescaled in order to withdraw the
dependence on $\Omega$. We rescale time by a factor $\Omega$ [ we divide
Hamiltonian~(\ref{eqn:ham}) by $\Omega$]. We rescale the actions $J$ and $L$
by a factor $\lambda=\Omega^{1/3}$, i.e.\ we replace $H(J,L,\theta,\psi)$ by 
$\lambda H(J/\lambda,L/\lambda,\theta,\psi)$. We notice that this rescaling
does not modify the eccentricity $e$. The resulting Hamiltonian becomes~: 
\begin{equation}  \label{eqn:hamham}
H=H_0+F^{\prime}J^2 \sum_{n=-\infty}^{+\infty} V_n\cos (n\theta+\psi),
\end{equation}
where 
\begin{equation}  \label{eqn:sint}
H_0=-\frac{1}{2J^2}-L +J^4 \frac{\omega_c^{\prime 2}}{8}%
\sum_{n,m=-\infty}^{+\infty} V_m V_{m-n} \cos n\theta .
\end{equation}
and $F^{\prime}=F\Omega^{-4/3}$ is the rescaled amplitude of the field and $
\omega^{\prime}_c= \omega_c/\Omega$. In what follows we assume that $
\Omega=1$. Furthermore, for simplicity we assume that $e$ is a parameter of
the system equal to the initial eccentricity of the orbit in the Keplerian
problem ($\omega_c=0$ and $F=0$). \\
In this paper, we consider the high-scaled frequency regime for co-rotating 
orbits~: $\Omega/\omega_K >1$ where the Kepler frequency $\omega_K$ is equal to 
$1/J^3$, i.e., we consider part of phase space corresponding to $0<J<1$.

\subsection{Study of the integrable part of Hamiltonian~(\ref{eqn:ham})}

Several key features of the dynamics of Hamiltonian~(\ref{eqn:ham}) can be 
understood by looking at the integrable part. We consider the mean value with 
respect to $\theta$ of the integrable Hamiltonian~(\ref{eqn:sint})~:
\begin{equation}  \label{eqn:Hint}
\tilde{H}_0=-\frac{1}{2J^2}-L +J^4 \frac{\omega_c^{2}}{8}\Vert V\Vert ^2,
\end{equation}
where $\Vert V\Vert^2=\sum_{n=-\infty}^{+\infty} V_n^2$, or another way for
considering $\tilde{H_0}$ is to assume that $J^4\frac{\omega_c^2}{8} 
\sum_{n\not= 0, m} V_mV_{m-n}\cos n\theta$ is part of the perturbation 
of Hamiltonian~(\ref{eqn:hamham}). One of the main features of Hamiltonian 
$\tilde{H_0}$ is that it does not satisfy the standard twist condition for 
$\omega_c\not= 0$. Since the Hessian of this Hamiltonian is
\[
\frac{\partial^2 \tilde{H_0}}{\partial J^2}=\frac{3}{J^4}\left(\frac{1}{2}
J^6\omega_c^2 \Vert V\Vert ^2-1\right), 
\]
the phase space is divided into two main parts~: a positive twist region where $
\partial^2\tilde{H_0}/\partial J^2>0$ for
$J\geq (\sqrt{2}/\omega_c\Vert V\Vert)^{1/3}$, and a negative one where $\partial^2
\tilde{H_0}/\partial J^2<0$ for $J \leq (\sqrt{2}/\omega_c\Vert V\Vert)^{1/3}$.
Both regions are separated by a twistless
region ($\partial^2\tilde{H_0}/\partial J^2=0$). We notice that the
singularity at $J=0$ is located inside the negative twist region, that the energy in the negative
twist region is negative, and that
the positive twist and twistless regions do not exist in the absence of
magnetic field. Furthermore, the size of the negative twist region decreases like
$\omega_c^{-1/3}$ as one increases the magnetic field $\omega_c$, i.e., it 
shrinks to the singularity of the Hamiltonian ($J=0$). \\

\begin{figure}
\centerline{\includegraphics*[scale=0.4]{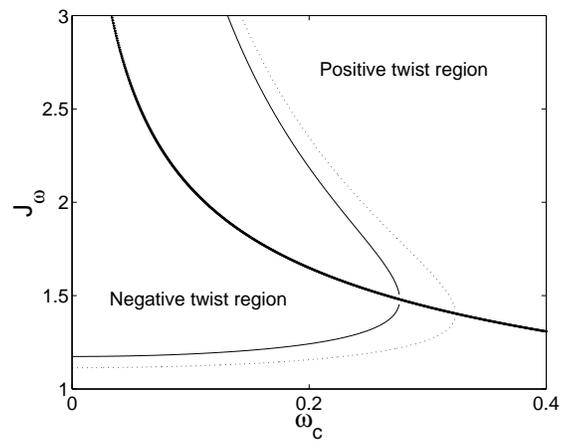}}
 \caption{\label{fig:Joms} Position of the invariant tori with frequency 
$(\sqrt{5}-1)/2$  (continuous curve) and $(5+\sqrt{5})/10$ (dashed curve)
 as a function of $\omega_c$ for the integrable Hamiltonian~(\ref{eqn:Hint}) for $e=1$.
 The strong continuous curve is the location of the twistless region.}
\end{figure}

The phase space of $\tilde{H_0}$, as well as the one of Hamiltonian~(\ref{eqn:sint}),
is foliated by invariant tori, the main difference being that the invariant tori for $\tilde{H_0}$ are flat in these
coordinates. We consider
a motion with frequency $\omega$, i.e.\ such that the dynamics is $\theta(t)=\omega t +\theta(0) \mod 2\pi$ and
$J(t)=J(0)$. The associated invariant torus is located
at $J_{\omega}$ such that the frequency $\frac{\partial \tilde{H}_0}{
\partial J}$ at $J=J_{\omega}$ is equal to $\omega$. The equation
determining $J_{\omega}$ is~: 
\begin{equation}
\frac{\omega_c^2}{2} \Vert V \Vert^2 J_{\omega}^6 -\omega J^3_{\omega}+1=0.
\end{equation}
There are two real positive solutions of this equation~: 
\[
J_{\omega}^{\pm}=\left(\frac{\omega\pm \sqrt{\omega^2-2\omega_c^2\Vert
V\Vert ^2}}{\omega_c^2\Vert V\Vert^2}\right)^{1/3}. 
\]
The condition of existence of an invariant torus with frequency $\omega$
for $\tilde{H_0}$ is $\omega \geq \sqrt{2} \omega_c \Vert V\Vert$. There are two
invariant tori with frequency $\omega$: one located at $J_\omega^-$ in the
negative twist region is a continuation of the torus with frequency $\omega$
in the absence of magnetic field since $\lim_{\omega_c\to 0} J_{\omega}^-
=\omega^{-1/3}$; the other torus located at $J_\omega^+$ in the positive
twist region is created far from the singularity $J=0$ as soon as the field
is non-zero. Figure~\ref{fig:Joms} depicts the
position of these tori. We notice that as soon as $\omega_c\not= 0$
there is creation of a set of invariant tori far from $J=0$ in
the positive twist region.\newline
If we increase $\omega_c$, the
position of the positive twist torus decreases whereas the position of
the negative twist torus increases. Both tori collide at
$\omega_c=\omega/\sqrt{2}\Vert V\Vert$ to a twistless invariant torus
(of the same frequency $\omega$) located at $ (\omega/2)^{-1/3}$. The
approximate location of this twistless torus as a function of
$\omega_c$ (for fixed parameter $e$) is plotted in Fig.~\ref{fig:Joms}.
As one increases $\omega_c$, a large portion of the invariant tori with 
$\omega\in [0,1]$ disappears.

{\em Remark : First Order Delaunay normalization}~\cite{lanc97} \newline
Averaging Hamiltonian~(\ref{eqn:hamham}) over $\theta$ gives 
\[
\langle H\rangle_{\theta} = -\frac{1}{2J^2}-L +J^4 \frac{\omega_c^2}{8}
\Vert V\Vert^2 -\frac{3e}{2}J^2 F\cos\psi. 
\]
Using the expansion of $\Vert V\Vert^2$ to the second order of $e$, $\Vert
V\Vert^2\approx 1+\frac{3e^2}{2}$, and the fact that the previous
Hamiltonian does not depend on $\theta$ (thus $J$ is constant), the
Hamiltonian reduces to
\[
{\cal K}=-L+\frac{3\omega_c^2}{16}e^2J^4-\frac{3}{2}eJ^2F\cos\psi, 
\]
which is the Hamiltonian studied in Ref.~\cite{lanc97} to find bifurcation
of equilibrium points.

\subsection{Primary main resonances and Chirikov resonance overlap}
\label{sec:chir}

\begin{figure}
\centerline{\includegraphics*[scale=0.4]{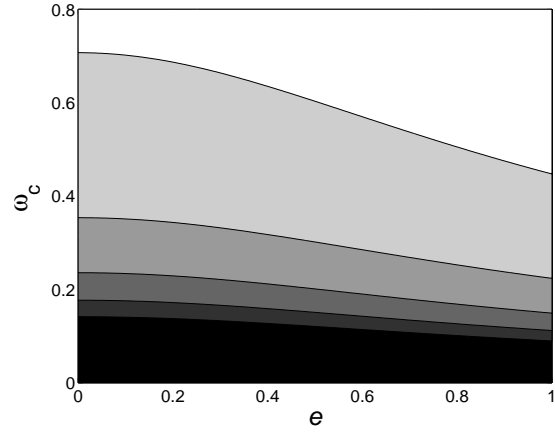}}
 \caption{\label{fig:omse} Existence of primary resonances ($m=1,\ldots,5$)
 in the plane of parameters $e$-$\omega_c$. The light gray part is the
domain of  existence of only $m=1$, and in the black part, all the
five first resonances exist.} 
\end{figure}

The positive and negative twist regions have their own sets of primary
resonances given by Hamiltonian~(\ref{eqn:hamham}). The approximate
locations of these primary resonances $m$:1, denoted $J_m^{\pm}$, are
obtained by the condition $m\dot{\theta}+\dot{\psi}\approx 0$. There
are two such resonances located approximately at \begin{equation}
\label{eqn:posres} J^{\pm}_m=\left( \frac{1\pm
\sqrt{1-2m^2\omega_c^2\Vert V\Vert^2}}{ m\omega_c^2\Vert V
\Vert^2}\right)^{1/3}. \end{equation}
The resonance located at $J_m^-$ is the continuation of the resonance in the
case $\omega_c=0$. The condition of existence of these resonances is $m\sqrt{
2}\omega_c\Vert V\Vert<1$. Similarly to collisions of invariant tori,
collisions of periodic orbits occur when increasing $\omega_c$. Figure~\ref
{fig:omse} depicts the different domains of existence of real $J_m$ for
first primary resonances $(m=1,\ldots,5)$ in the plane of parameters $e$-$
\omega_c$. We notice that the most relevant parameter in this problem is the
magnetic field. The variations of the dynamics induced by the eccentricity
$e$ are smooth.

Between two consecutive resonances $m$:1 and $m$+1:1 (if they exist), regular and
chaotic motions occur for a typical value of the field $F$. In order
to estimate the chaos threshold between these resonances, i.e.\ the
value for which there is no longer any rotational invariant torus
acting like a barrier in phase space, Chirikov overlap criterion
provides an upper bound but usually quite far from realistic values
(obtained by numerical integration for instance). For
quantitatively accurate thresholds, a modified criterion is applied~:
the 2/3-rule criterion.

In order to compute the resonance overlap value of the field $F$ between $m$:1
and $m$+1:1 primary resonances,
we follow the procedure described in Ref.~\cite{meer82}. First, we
change the frame to a rotating one at the phase velocity of resonance
$m$:1. We apply the following canonical change of variables~:
$(\boldsymbol{A}^{\prime},\boldsymbol{\varphi}^{\prime})=(
\tilde{N}^{-1}\boldsymbol{A},N\boldsymbol{\varphi})$ where
$\u{A}=(J,L)$ and $\u{\varphi}=(\theta,\psi)$ and \[ N=\left(
\begin{array}{cc}
m & 1 \\
0 & 1
\end{array}
\right),
\]
and $\tilde{N}$ denotes the transposed matrix of $N$. Hamiltonian~(\ref
{eqn:hamham}) is mapped into
\begin{eqnarray*}
H=&&-\frac{1}{2m^2J^{\prime
2}}-J^{\prime}-L^{\prime}+\frac{\omega_c^2}{8}m^4J^{\prime
4}\sum_{n,n^{\prime}} V_{n^{\prime}}V_{n^{\prime}-n}
\cos[n(\theta'-\psi')/m] \\ &&+Fm^2J^{\prime 2}\sum_n
V_n\cos[(n\theta'+(m-n)\psi')/m]. \end{eqnarray*} By averaging over
the fast variable $\psi'$, the Hamiltonian becomes \[
H=-\frac{1}{2m^2J^{\prime 2}}-J^{\prime}+\frac{\omega_c^2}{8}m^4
J^{\prime 4}\Vert V\Vert^2+Fm^2 V_m J^{\prime 2}\cos \theta'. \]
The resulting Hamiltonian is integrable and one can compute the width of the
resonance $m$:1 following Ref.~\cite{meer82}. We expand the previous
Hamiltonian around $J'_m=J_m/m$ and keep only the quadratic part in
$\Delta J'=J'-J'_m$ and the constant term in the action $\Delta J'$
proportional to $\cos \theta'$~:
\[
H=-\frac{3m^2}{2J_m^4}\left(1-\frac{1}{2}\omega_c^2\Vert
V\Vert^2 J_m^6\right) \Delta J^{\prime 2}
+FJ_m^2V_m \cos \theta'.
\]
We rescale energy by a factor $-3m^2J_m^{-4}(1-\omega_c^2\Vert
V\Vert^2J_m^6/2)$~: \[
H=\frac{1}{2}\Delta J^{\prime 2}
-\frac{FJ_m^6V_m}{3m^2\left(1-\frac{1}{2}\omega_c^2\Vert V\Vert^2
J_m^6\right)}\cos \theta'. \]
Depending on the positive or negative twist region, this rescaling is
positive or negative respectively.

The width of the resonance $m$:1 in the
variables $\Delta J= m\Delta J'$ is given by~:
\[ \Delta_m=4 J_m^3\sqrt{\frac{F
V_m(e)}{3}}\left| 2-\frac{J_m^3}{m}\right|^{-1/2}, \]
since $\omega_c^2\Vert V\Vert^2 J_m^6/2=-1+J_m^3/m$.
The 2/3-rule criterion for the
critical threshold for the overlap between resonance $m$:1 and $m$+1:1
is reached when the distance between two neighboring primary
resonances is equal (up to a factor $2/3$) to the sum of the
half-widths of these resonances~: \[
\frac{2}{3}\vert J_{m+1}-J_m\vert =\frac{1}{2}(\Delta_{m}+\Delta_{m+1}).
\]
The critical threshold is given by~:
\begin{equation}  \label{eqn:critchir}
F_m(e,\omega_c)=\frac{(J_{m+1}-J_m)^2}{3\left(
J_m^3\sqrt{V_m}\left| 2-\frac{J_m^3}{m}\right|^{-1/2}+J_{m+1}^3
\sqrt{V_{m+1}}\left| 2-\frac{J_{m+1}^3}{m+1}\right|^{-1/2} \right)^2},
\end{equation} where $J_m$ stands for either $J_m^+$ or $J_m^-$ given
by Eq.~(\ref {eqn:posres}). Therefore we obtain two critical
couplings~: one in the positive twist region, $F_m^+(e,\omega_c)$,
and one in the negative twist region, $F_m^-(e,\omega_c)$.
We notice that for $\omega_c=0$, since $J_m^-=m^{1/3}$, we obtain the
formula of Ref.~\cite{sach97} for the chaos threshold in the CP
problem. For $\omega_c=1/[\sqrt{2}\Vert V\Vert (m+1)]$, the
threshold $F_m(e,\omega_c)$ vanishes. This case corresponds to the
collision of the resonances $m$+1:1 (the positive and negative twist
ones). Therefore, the Chirikov criterion is valid only for
$\omega\leq 1/[\sqrt{2}\Vert V\Vert (m+1)]$, i.e.\ when the two primary
resonances $m$:1 and $m$+1:1 exist in the system.\\

We use this criterion in order to study the stability in the different
regions of phase space as a function of the magnetic field $\omega_c$ (for small 
values of the field $\omega_c$) and the eccentricity of the initial orbit $e$. 
However, since this criterion is purely empirical, we use another method to 
validate or refine the results~: we use the renormalization method which has
proved to be a very powerful and accurate method for determining chaos
thresholds in this type of models~\cite{chan01R,chan01a,chan02a,chan00d}.  We 
compare the results given by both methods in the region where the criterion 
applies and we use the renormalization map to compute chaos thresholds when
Eq.~(\ref{eqn:critchir}) does not apply (when one of the main primary resonance
$m$:1 has disappeared).

\subsection{Renormalization method}

\label{sec:reno}

The Chirikov resonance overlap criterion gives us a value for the
chaos threshold between two neighboring primary resonances. This value
aims at approximating the value of the field $F$ for which there is no
longer any barrier in phase space, and for which large-scale diffusion
of trajectories occur between these resonances. The renormalization method
gives a more local and more accurate information. This method
computes the threshold of break-up of an invariant torus with a given
frequency $\omega$. Then by varying $\omega$, we obtain the global
information on the transition to chaos.

\subsubsection{Expansion of the Hamiltonian around a torus}

In order to apply the renormalization transformation as defined in
Refs.~\cite {chan99c,chan00a} for a given torus with frequency
$\omega$, we expand Hamiltonian~(\ref{eqn:ham}) in Taylor series in
the action $J$ around $ J=J_{\omega}$.
\[
-\frac{1}{2 J^2}=\sum_{k=0}^{+\infty} (-1)^{k+1}\frac{k+1}{2 J_{\omega}^{k+2}
} \Delta J ^k, 
\]
where $\Delta J = J-J_{\omega}$. The meanvalue of the quadratic term in $
\Delta J$ of $H_0$ is equal to $\frac{3(\omega J_{\omega}^3-2)}{2
J_{\omega}^4} \Delta J ^2$. We rescale the action variables such that this
quadratic term is equal to 1/2, i.e.\ we replace $H(\Delta J,L,\theta,\psi)$
by $\lambda H(\Delta J/\lambda,L/\lambda,\theta,\psi)$ with $
\lambda=3(\omega J_{\omega}^3-2)/J_{\omega}^4$. We notice that this
rescaling can be done except for two cases $J_{\omega}=(\omega/2)^{-1/3}$
which is the twistless case, and for $J_{\omega}=0$ which is the
singularity. Therefore, as it is defined in Refs.~\cite{chan01R,chan99c,chan00a},
the renormalization cannot be applied in the twistless region.

The Hamiltonian~(\ref{eqn:ham}) becomes~: 
\begin{equation}  \label{eqn:hamresc}
H_F=\omega \Delta J -L +\sum_{k\geq 2}^{+\infty} H_{k,0} \Delta J^k
+\sum_{k=0}^4 \sum_{n >0} H_{k,n} \Delta J^k \cos n\theta +F\sum_{k=0}^2
\sum_{n=-\infty}^{+\infty} V_{k,n}\Delta J^k \cos (n\theta+\psi),
\end{equation}
where 
\begin{eqnarray*}
&& H_{2,0}=\frac{1}{2}, \\
&& H_{3,0}=\frac{J_{\omega}^3(\omega J_{\omega}^3+1)}{9(2-\omega
J_{\omega}^3)^2}, \\
&& H_{4,0}=\frac{J_{\omega}^6(11-\omega J_{\omega}^3)}{108(2-\omega
J_{\omega}^3)^3}, \\
&& H_{k,0}=\frac{(k+1)J_{\omega}^{3k-6}}{2 \cdot 3^{k-1}(2-\omega
J_{\omega}^3)^{k-1}} \qquad \mbox{ for } k\geq 5, \\
&& H_{0,n}=-\frac{3}{4}\omega_c^2(2-\omega
J_{\omega}^3)\sum_{m=-\infty}^{+\infty} V_m V_{m-n}, \\
&& H_{1,n}=\omega_c^2 J_{\omega}^3 \sum_{m=-\infty}^{+\infty} V_m V_{m-n}, \\
&& H_{2,n}=-\frac{\omega J_{\omega}^3 -1}{2-\omega J_{\omega}^3} \cdot \frac{
\sum_{m=-\infty}^{+\infty} V_m V_{m-n}}{\Vert V \Vert ^2}, \\
&& H_{3,n}=\frac{2J_{\omega}^3 (\omega J_{\omega}^3 -1)}{9(2-\omega
J_{\omega}^3)^2} \cdot \frac{ \sum_{m=-\infty}^{+\infty} V_m V_{m-n}}{\Vert
V \Vert ^2}, \\
&& H_{4,n}=-\frac{J_{\omega}^6 (\omega J_{\omega}^3 -1)}{54(2-\omega
J_{\omega}^3)^3} \cdot \frac{ \sum_{m=-\infty}^{+\infty} V_m V_{m-n}}{\Vert
V \Vert ^2}, \\
&& V_{0,n}=-3 V_n \frac{2-\omega J_{\omega}^3}{J_{\omega}^2}, \\
&& V_{1,n}=2 J_{\omega} V_n, \\
&& V_{2,n}=- V_n \frac{J_{\omega}^4}{3(2-\omega J_{\omega}^3)},
\end{eqnarray*}

for $n\in {\Bbb Z}$. The resulting Hamiltonian is of the form~: 
\begin{equation}  \label{eqn:form}
H_F(\boldsymbol{A},\boldsymbol{\varphi})=\boldsymbol{\omega}\cdot 
\boldsymbol{A} + V(\boldsymbol{\Omega}\cdot\boldsymbol{A},
\boldsymbol{\varphi}),
\end{equation}
where $\boldsymbol{\omega}=(\omega,-1)$ and $\boldsymbol{\Omega}=(1,0)$, and
the coordinates are $\boldsymbol{A}=(\Delta J,L)$ and $\boldsymbol{\varphi}
=(\theta,\psi)$. For small $F$, the Kolmogorov-Arnold-Moser (KAM)
theorem states that if $\omega$ satisfies a Diophantine condition, the
invariant torus with frequency $ \omega $ will persist. In fact, the
picture that emerges from numerical simulations is that if $F$ (in
absolute value) is smaller than some critical threshold $F_c(\omega)$
then Hamiltonian~(\ref{eqn:ham}) [or equivalently (\ref{eqn:hamresc})]
has an invariant torus with frequency $\omega$. If $F$ is larger than
this critical threshold, the torus is broken. In order to compute
numerically the critical function $\omega\mapsto F_c(\omega)$, we use
the renormalization method described briefly below.

\subsubsection{Renormalization method}

Without restriction (up to a rescaling of time), we assume that
$\omega\in [0,1]$. The renormalization relies upon the
continued fraction expansion of the frequency $\omega$ of the torus:
$$ \omega=\frac{1}{a_0+\frac{1}{a_1+\cdots}} \equiv [a_0,a_1,\ldots].$$

This transformation will act within a space of
 Hamiltonians $H$ of the form
\begin{equation}
\label{eqn:HRG}
H(\u{A},\u{\varphi})=\u{\omega}\cdot\u{A}+V(\u{\Omega}
\cdot\u{A},\u{\varphi}),
\end{equation}
where $\u{\Omega}=(1,\alpha)$ is a vector not parallel to the
frequency vector $\u{\omega}=(\omega,-1)$. We assume that 
Hamiltonian~(\ref{eqn:HRG}) satisfies a non-degeneracy condition~:
 If
we expand $V$ into
$$
V(\u{\Omega}\cdot\u{A},\u{\varphi})=\sum_{\scriptstyle \u{\nu} \in
{\mathbb{Z}}^2 \atop
k\geq 0} V^{(k)}_{\u{\nu}}
(\u{\Omega}\cdot\u{A})^k e^{i\u{\nu}\cdot\u{\varphi}},
$$
we assume that $V_{\boldsymbol{0}}^{(2)}$ is non-zero. This
restriction means that we cannot explore the twistless region
by the present renormalization transformation.\newline
The
transformation contains essentially two parts~: a rescaling and an
elimination procedure~\cite{koch99}.

 {\bf (1)} \textit{Rescaling}~: The first part of the transformation is
 composed by a shift of the resonances, a rescaling of time and a
 rescaling of the actions. It acts on a Hamiltonian $H$ as $H'=H\circ
 {\mathcal T}$ (see Ref.~\cite{chan00a} for details)~:
\begin{equation}
H'(\u{A},\u{\varphi})=\lambda\omega^{-1} H\left(-\frac{1}{\lambda}N_{a}\u{A},
-N_{a}^{-1}\u{\varphi}\right),
\end{equation}
with
\begin{equation}
\label{eqn:rescaling}
\lambda=2\omega^{-1} (a+\alpha)^2 V^{(2)}_{\u{0}},
\end{equation}
and
\begin{equation}
  \label{eq:matrix2d}
  N_{a}=\left(\begin{array}{cc} a & 1\\ 1 & 0 \end{array}\right),
\end{equation}
and $a$ is the integer part of $\omega^{-1}$ (the first entry in the
continued fraction expansion). This change of coordinates is a
generalized (far from identity) canonical transformation and the
rescaling $\lambda$ is chosen to ensure a normalization
condition (the quadratic term in the actions has a mean value equal to
1/2). For $H$ given by Eq.\ (\ref{eqn:HRG}), this expression becomes
\begin{equation} H'(\u{A},\u{\varphi})=\u{\omega}'\cdot\u{A}+
\sum_{\u{\nu},k} V^{\prime (k)}_{\u{\nu}}
(\u{\Omega}'\cdot\u{A})^k e^{i\u{\nu}\cdot\u{\varphi}},
\end{equation}
where
\begin{eqnarray}
  \label{eq:renexp}
  && \u{\omega}'=(\omega',-1) \; \mbox{ with }
  \omega'=\omega^{-1}-a,\\
  && \u{\Omega}'=(1,\alpha) \; \mbox{ with } \alpha'=1/(a+\alpha),\\
  && V^{\prime (k)}_{\u{\nu}}=r_k V_{-N\u{\nu}}^{(k)} \; \mbox{ with }
  r_k=(-1)^k 2^{1-k}\omega^{k-2}(a+\alpha)^{2-k} \left(
  V^{(2)}_{\u{0}}\right) ^{1-k}.
\end{eqnarray}
We notice that
the frequency of the torus is changed according to the Gauss map~:
\begin{equation}
\label{eqn:gauss}
\omega \mapsto \omega'=\omega^{-1}-\left[ \omega^{-1}\right],
\end{equation}
where $\left[\omega^{-1} \right]$ denotes the integer part of $\omega^{-1}$.
Expressed in terms of
the continued fraction expansion, it corresponds to a shift to the
left of the entries
$$
\omega=[a_0,a_1,a_2,\ldots]\mapsto \omega'=[a_1,a_2,a_3,\ldots].
$$

{\bf (2)} \textit{Elimination}~: The second step is a (connected to
identity) canonical transformation ${\mathcal U}$ that eliminates the
non-resonant modes of the perturbation in $H'$.
Following Ref.~\cite{chan00a}, we consider the
set $I^- \subset {\mathbb{Z}}^2$ of non-resonant modes as the set of integer
vectors $\u{\nu}=(\nu_1,\nu_2)$ such that $|\nu_2|> |\nu_1|$.
The canonical transformation ${\mathcal U}$ is such that $H''=H'\circ
{\mathcal U}$ does not have any non-resonant mode, i.e.\ it is
defined by the following equation~:
\begin{equation}
  \label{eq:proj}
  {\mathbb{I}}^-(H'\circ {\mathcal U})=0,
\end{equation}
where ${\mathbb{I}}^-$ is the projection operator on the non-resonant
modes; it acts on a Hamiltonian~(\ref{eqn:HRG}) as~:
$$
{\mathbb{I}}^- H=\sum_{\scriptstyle \u{\nu}\in I^- \atop k\geq 0}
V_{\u{\nu}}^{(k)}
(\u{\Omega}\cdot\u{A})^k e^{i\u{\nu}\cdot\u{\varphi}}.
$$
We solve Eq.~(\ref{eq:proj}) by a Newton method following
Refs.~\cite{chan98b,chan00a}.

Thus the renormalization acts on $H$ for a torus with frequency
$\omega$ as $H''={\mathcal R}(H)=H\circ {\mathcal T} \circ{\mathcal
U}$ for a torus with frequency $\omega'$.  \\
The critical thresholds are obtained by iterating the renormalization
transformation $\mathcal{R}$.
The main conjecture
of the renormalization approach is that if the
torus exists for a given Hamiltonian $H$, the iterates ${\mathcal{R}}^n H$
of the renormalization map acting
on $H$
converge to some integrable Hamiltonian $H_0$. This
conjecture is supported by analytical results in the perturbative
regime~\cite{koch99,koch99b}, and by numerical 
results~\cite{chan00d,chan01a,chan01R}. For a one-parameter family of 
Hamiltonians $\{H_F\}$, the critical amplitude of the perturbation
$F_c(\omega)$ is determined by the following conditions~:
\begin{eqnarray} && {\mathcal R}^nH_{F} \underset{n\to \infty}{\to}
H_0(\u{A}) =\u{\omega}\cdot\u{A}+ \frac{1}{2}(\u{\Omega}\cdot\u{A})^2
\quad \mbox{ for } F<F_c(\omega), \label{eqn:Adef1}\\ && {\mathcal
R}^nH_{F} \underset{ n\to \infty}{\to} \infty \quad \mbox{ for }
F>F_c(\omega).\label{eqn:Adef2} \end{eqnarray}
The critical threshold $F_c(\omega)$ is determined by a bisection
procedure.\\
In order to obtain the value $F_m$ of chaos threshold between
resonances $m$:1 ans $m$+1:1, we vary $\omega$ between $1/(m+1)$ and
$1/m$. The critical threshold is given by $F_m=\max_{\omega\in
[1/(m+1),1/m]} F_c(\omega)$.

\section{Numerical computation of chaos thresholds}

\label{sec:resu}

\subsection{Chaos thresholds in the negative twist region}
\label{sec:neg}

\begin{figure}
\centerline{\includegraphics*[scale=0.4]{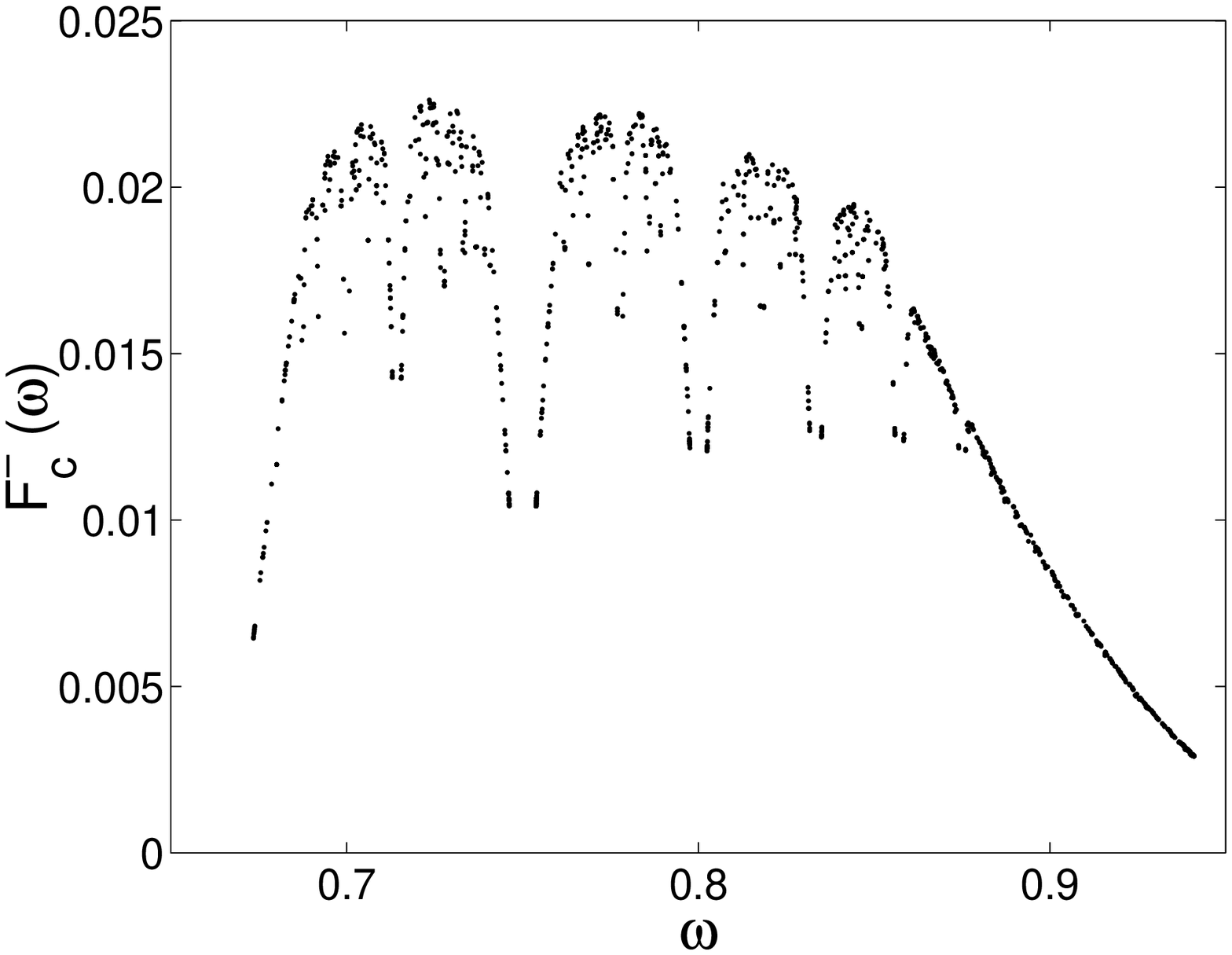}}
 \caption{\label{fig:fomn} Critical function $F_c^-(\omega)$ in the  negative 
twist region, obtained by the renormalization method, for $\omega_c=0.3$ and for 
$e=1$.} 
\end{figure}

Figure~\ref{fig:fomn} represents a typical plot of the critical
function $\omega\mapsto F_c^-(\omega)$ in the negative twist region
obtained by the renormalization
method for $e=1$ and $\omega_c=0.3$. This function vanishes at
(at least) all rational values of the frequency $\omega$ since all tori with rational
frequency are broken as soon as the field is turned on.
The condition of existence of a
torus with frequency $\omega$ obtained from the integrable case~(\ref{eqn:sint})
is $\omega >
\sqrt{2}\omega_c\Vert V\Vert$, which is in that case $\omega > 0.67$.
We expect the tori with frequency belonging to [0,0.67] to be
broken
by collision with invariant tori in the positive twist region
before this value of the field $\omega_c$, as it is the case in
Hamiltonian~(\ref{eqn:sint}).

From Fig.~\ref{fig:fomn}, we deduce that for $F>F_c=0.023$, no
invariant tori are left in this region of phase space.
The frequency of the last invariant torus is equal to
$(\gamma+1)/(2\gamma+1)\approx 0.7236$ where
$\gamma=(\sqrt{5}-1)/2$. This value of the frequency of
the last invariant torus varies with the parameters
$e$ and $\omega_c$ (for $\omega_c=0$, see Fig.~2 of Ref.~\cite{chan02a}). The 
main feature of this frequency is that it remains noble as the parameters
$e$ and $\omega_c$ are varied,
in the sense that the tail of the continued fraction expansion of this
frequency is a sequence of 1, or equivalently this frequency expresses
like $(a\gamma+b)/(c\gamma+d)$ where $a$, $b$, $c$, $d$ are integers
such that $ad-bc=\pm 1$.\\

\begin{figure}
\centerline{\includegraphics*[scale=0.4]{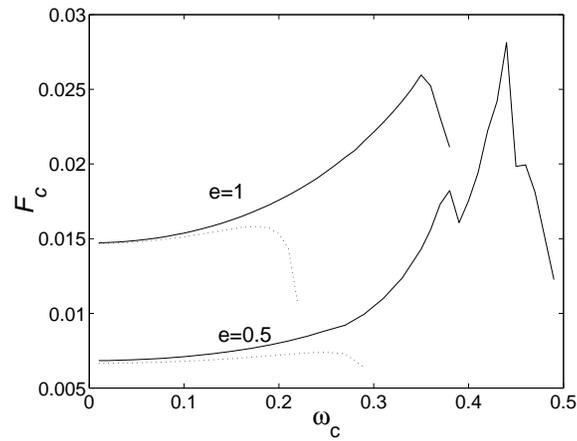}}
 \caption{\label{fig:renchirn} Chaos thresholds $F_1^-(\omega_c)$ between 
resonances 1:1 and 2:1 in the  negative twist region, obtained
 by the 2/3-rule criterion (dashed curves) and by the renormalization
method (continuous curves), as a function of $\omega_c$, for $e=0.5$ and $e=1$.}
\end{figure}

We compute chaos thresholds between two successive primary resonances
by two methods~: the 2/3-rule criterion and the renormalization map.
We compute the critical value of the field $F$ for which there
is no longer any invariant torus between resonances 1:1 and 2:1,
located at $ J_{1}^{-}$ and $J_{2}^{-}$ respectively, as a function of
the magnetic field $\omega_c$ and the parameter $e$ in the negative
twist region, i.e.\
$F_1^-(e,\omega_c)=\max_{\omega\in [1/2,1]} F_c^-(\omega,e,\omega_c)$.
Figure~\ref{fig:renchirn}
represents two computations~: for $e=0.5$ and for $e=1$.

For $\omega_c=0$, we obtain the values that have been obtained in
Ref.~\cite{chan02a}. The 2/3-rule criterion gives a very good
approximation for low values of the field $\omega_c$ (typically for
$\omega_c\in [0,0.15]$). For
small $\omega_c$, we develop the critical function
$F_1^-(\omega_c)$ given by Eq.~(\ref{eqn:critchir}).
The corrections to $ F_c^-$ due to the magnetic
field are of order $\omega_c^2$ since $J_n^-=n^{1/3} +O(\omega_c^2).$

We notice that at some value of $\omega_c$,
the curves $F_1^-(e,\omega_c)$ given by the Chirikov criterion decrease sharply 
to zero.
We have seen that the approximate condition of existence of the
resonance $m$:1, derived from the integrable case~(\ref{eqn:Hint}), is 
$\omega_c < 1/(\sqrt{2}m \Vert V \Vert)$. For $e=1$ and $m=2$, this condition is
$\omega_c<0.23$ (and for $e=0.5$ this condition becomes
$\omega_c<0.30$).  This means that the criterion derived in Sec.~\ref{sec:chir} 
is inapplicable for $\omega_c$ larger than these values ($J_2^-$ becomes 
complex).  One way of extending this
criterion for larger values of
$\omega_c$ would be to consider the overlap of higher order
resonances. From the renormalization map results, we see that after 2:1 
resonance disappear, there is an increase of stability which can be explained by 
the fact that the invariant tori in this region of phase space are no longer 
deformed by this neighboring resonance. We expect this region to be fully broken 
{\it before} 1:1 disappears (i.e.\ all the 
region between 1:1 and 2:1 primary resonances has disappeared), and this happens 
at $\omega_c\approx 0.45$ for $e=1$, and at $\omega_c\approx 0.6$ for
$e=0.5$. These results are consistent with the sharp decreases of $F_c$
found by renormalization on Fig.~\ref{fig:renchirn}.  \newline

\begin{figure}
\centerline{\includegraphics*[scale=0.4]{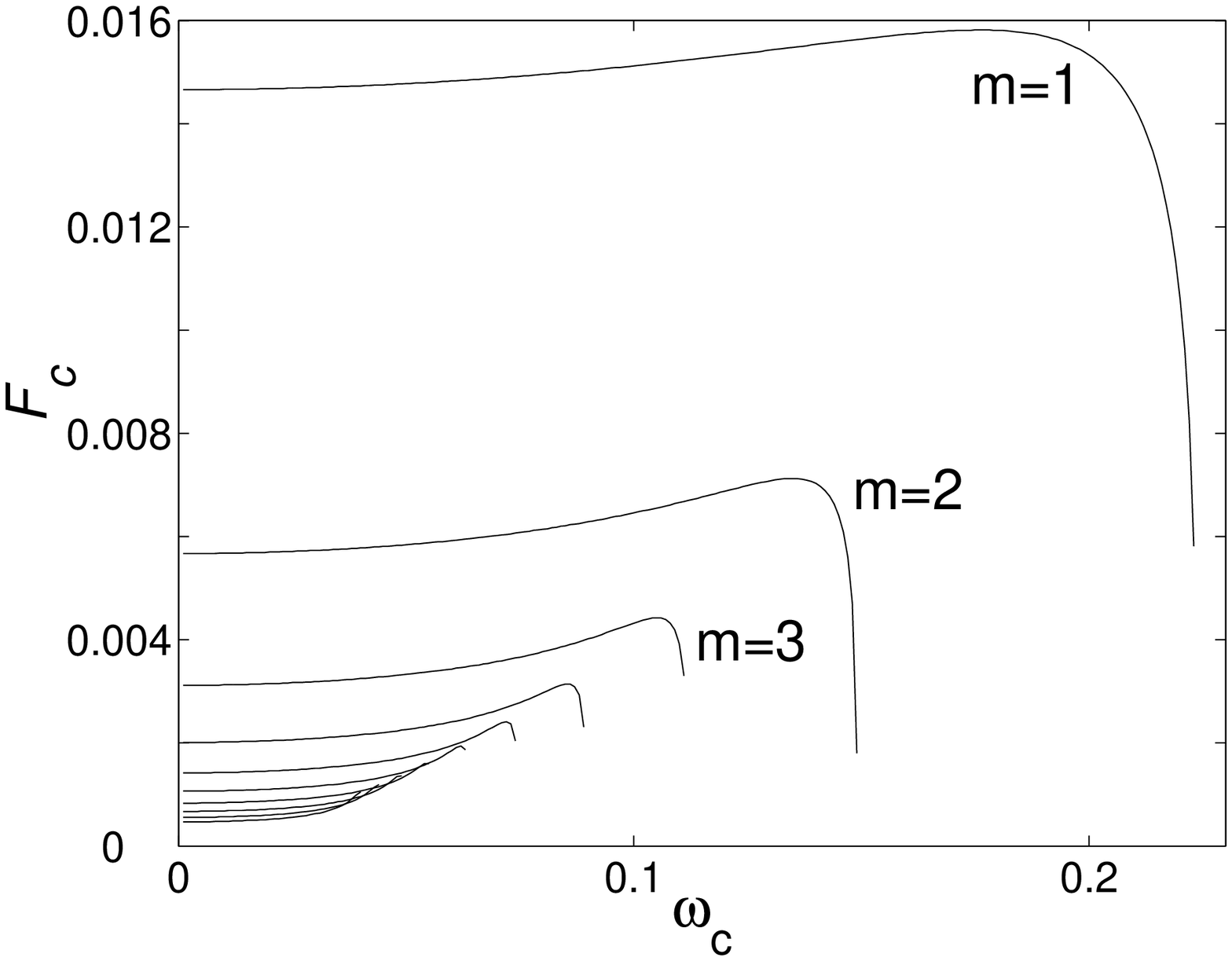}}
\caption{\label{fig:Epschirn} Resonance overlap values of 
$F_m^-(\protect\omega_c)$ between resonances $m$:1 and $m$+1:1 for $m=1,\ldots 
,10$ in the negative twist region for $e=1$.} 
\end{figure}

Furthermore, we investigate the chaos thresholds for the different
primary resonances $m$:1 by the Chirikov criterion.
Using Eq.~(\ref{eqn:critchir}), we compute the resonance overlap value of
$m$:1 and $m$+1:1 for increasing $m=1,\ldots,10$. For a fixed
eccentricity $e=1$, Fig.~\ref{fig:Epschirn} shows the resonance
overlap values as a function of the field $\omega_c$

From this figure it emerges that the dynamics in that region
is mainly insensitive to the magnetic field for low values of this field.
The sharp decreases of the curves $F_m^-$ are due to the disappearance
of one of the primary resonances from which the resonance overlap is computed.
At these values and for larger $\omega_c$, we expect stability enhancement
due to the disappearance of this primary resonance. For $\omega_c\geq 1/(\sqrt{2}m
\Vert V\Vert)$, the critical curve $F_m^-$ sharply decreases to zero
due to the disappearance of all the region between primary resonances $m$:1 and
$m$+1:1 into the twistless region. For instance, for $m=3$, the critical curve $F_3^-(\omega_c)$
increases slowly with $\omega_c$ for $\omega_c\in [0,0.11]$, then for
$\omega_c\in [0.11,0.15]$, we expect stabilization enhancement by the magnetic field,
and for $\omega_c$ larger than 0.15, the region of phase space between 3:1 and 4:1 resonances
disappears into the twistless region and $F_3^-(\omega_c)$ vanishes.

\begin{figure}
\centerline{\includegraphics*[scale=0.4]{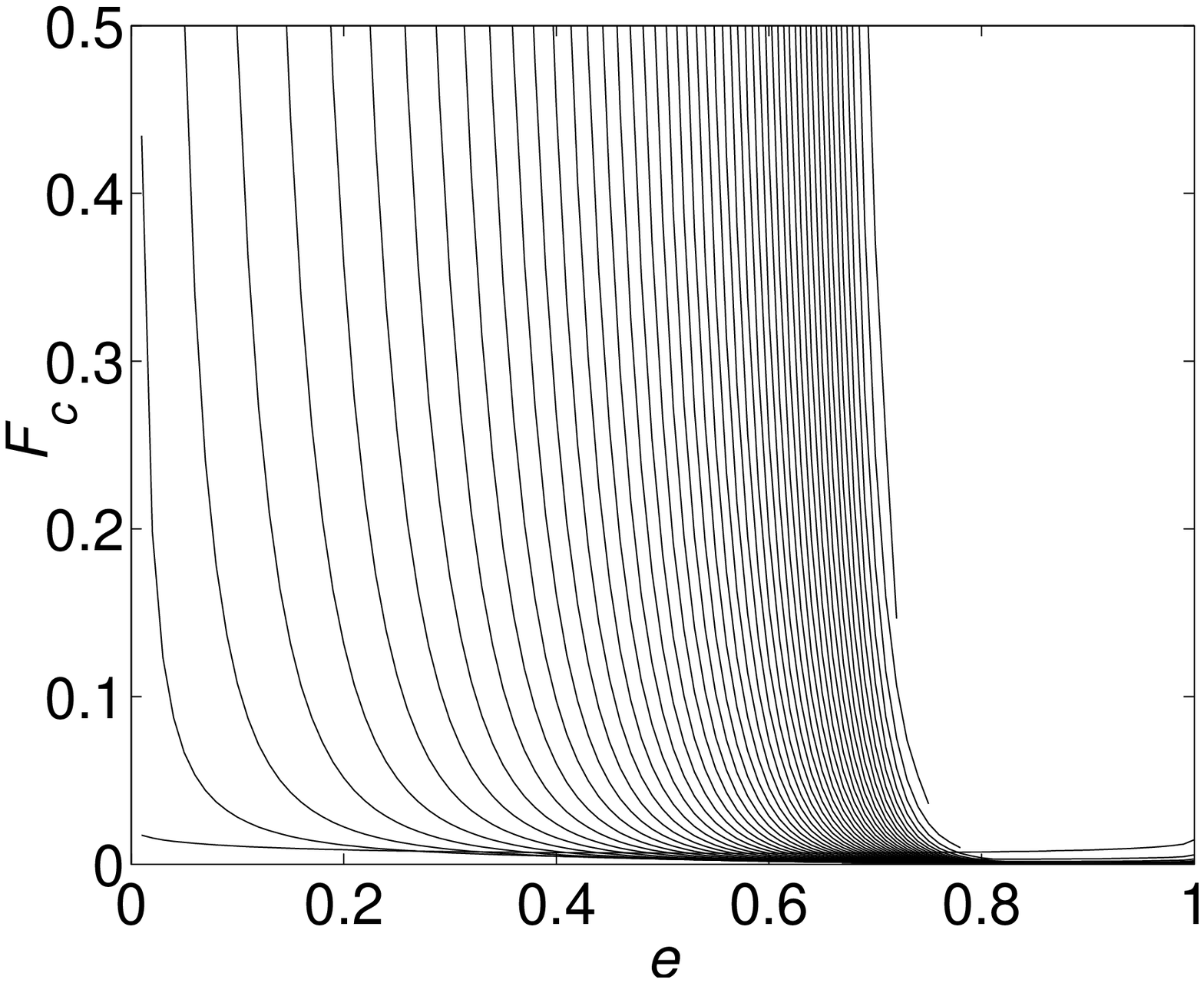}
\includegraphics*[scale=0.4]{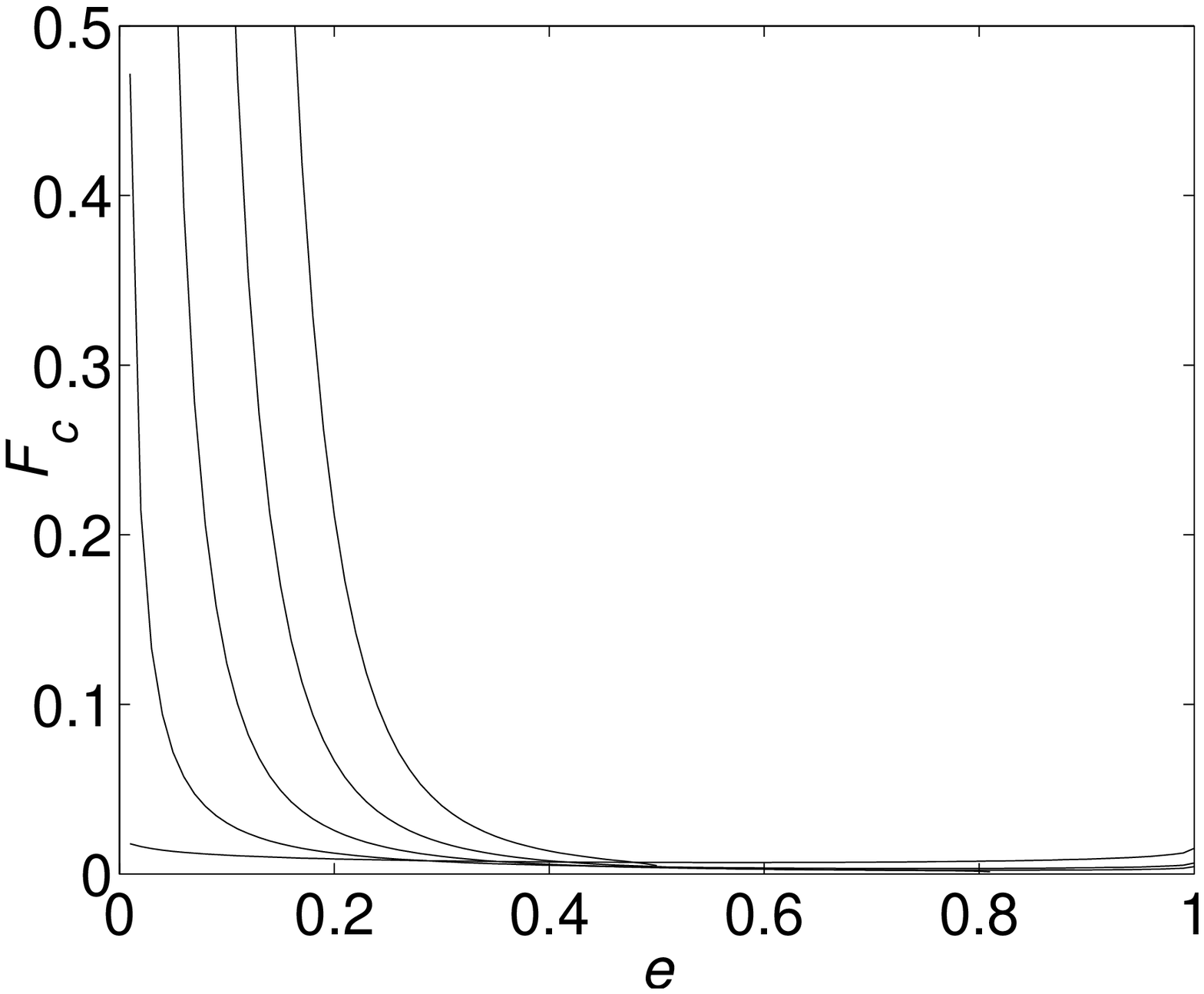}}           
\caption{\label{fig:fce} Resonance overlap values of $F_m^-(e)$ between
resonances $m$:1 and $m$+1:1 for $m=1,\ldots$ in the negative
twist region for $(a)$ $\omega_c=0.01$ and $(b)$ $\omega_c=0.1$.}
\end{figure}

The
region between resonances of low order ($m$ small) seem to be more stable for $e=1$.
However, this feature varies with the parameter $e$ as it has been
observed in Ref.~\cite{chan02a}. For $e\in [0,0.8]$ and for $\omega_c=0$, the 
regions near $m$:1 with large $m$ become very stable (and this stabilization is
increased by the field for low values of $\omega_c$) and the diffusion
of the trajectories is very limited. Therefore, the orbits with
eccentricity close to one are the easiest to diffuse in a broad range of phase space. In particular,
these orbits can ionize more easily than medium eccentricity ones.
This feature
is expected to hold for small $\omega_c$ (typically for $\omega_c\leq 0.05$)
even if the diffusion is now limited in the negative twist region since we expect the twistless
region to be very stable and act as a barrier in phase space~\cite{cast96}.
Figure~\ref{fig:fce} displays the values of $F_m^-(e)$ for
$e\in[0,1]$, and for two values of $\omega_c$~: $\omega_c=0.01$ and $\omega_c=0.1$.

For $\omega=0.1$, since a large portion of phase space between primary resonances with $m$ large
has disappeared into the twistless region,  the orbits with medium eccentricity become as easy to ionize as
the ones with eccentricity close to one in the remaining part of the negative twist region of phase space.
For
eccentricities close to one, the critical threshold is dominated by
the chaos threshold between resonances 1:1 and 2:1.

In summary, the effect of the magnetic
field $\omega_c$ is to stabilize the dynamics in the negative twist region by 
successively breaking up primary resonances. Therefore, between two succesive 
resonances with $m$ small, the effect of the magnetic field  is expected to be 
smooth in the sense that the critical thresholds 
are only slightly changed (increased) from the chaos thresholds in the absence 
of magnetic field up to some critical value of $\omega_c$ where
collisions of periodic orbits occur. Investigating this region by
Chirikov resonance overlap yields values which are qualitatively and
quantitatively accurate with respect to the renormalization results
for $\omega_c\in [0,0.1]$.\\
For small values of $\omega_c$, the qualitative behavior concerning diffusion of 
trajectories is expected to be the same as the one in the absence of magnetic 
field even if the diffusion coefficients may be smaller due to the limited 
region of phase space and the stability enhancement due to the magnetic field. 
Increasing $\omega_c$ makes the orbits with medium eccentricity easier to 
diffuse in a more and more reduced region of phase space.

\subsection{Chaos thresholds in the positive twist region}

The classical dynamics in the positive twist region of phase space is essential
for the stochastic ionization process since it contains the region where the 
action $J$ is large, anf for large $\omega_c$ this region is the predominant 
region since the negative twist region shrinks to the singularity $J=0$.

\begin{figure}
\centerline{\includegraphics*[scale=0.4]{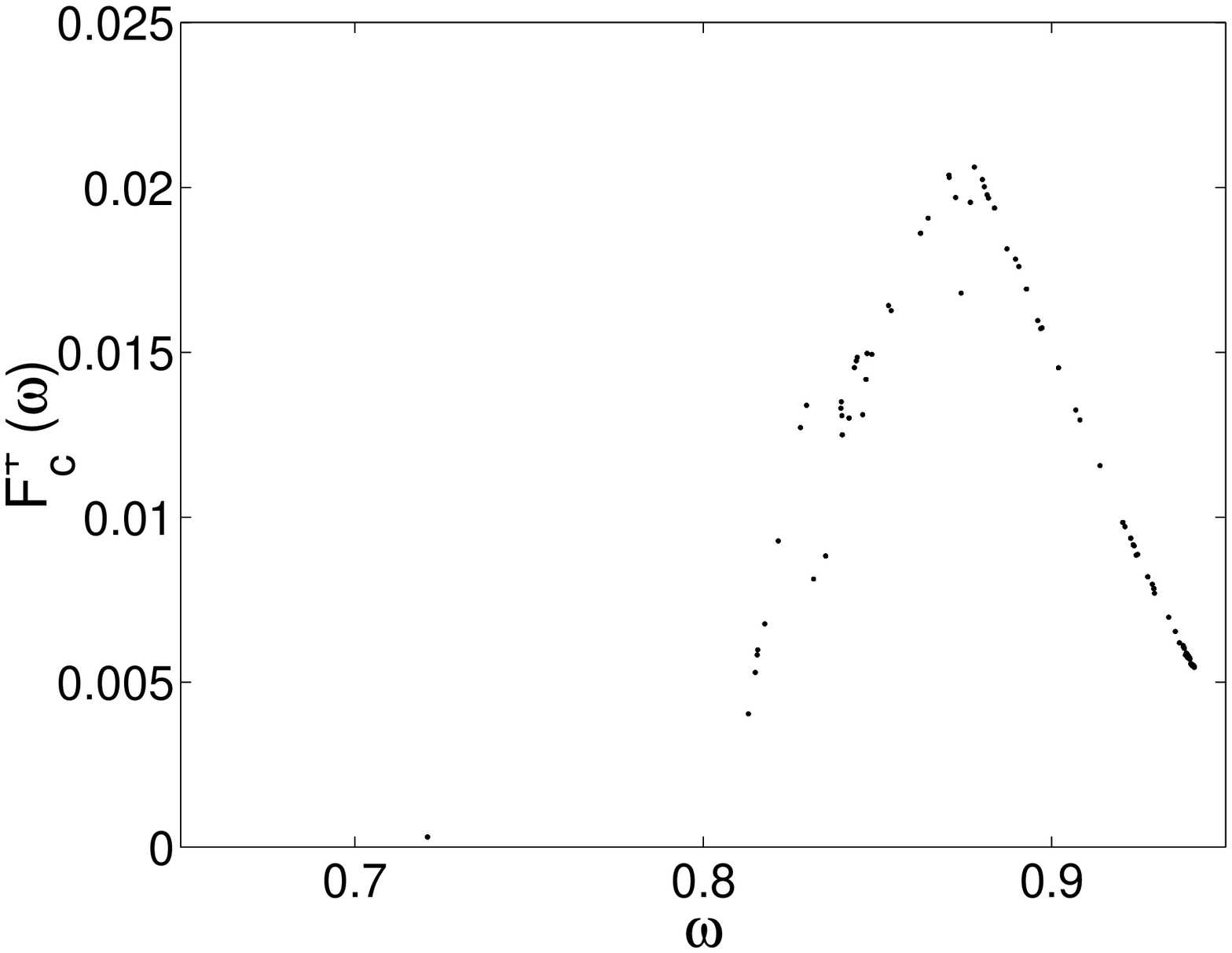}}
 \caption{\label{fig:fomp} Critical function $F_c^+(\omega)$ in the  positive 
twist  region, obtained by the renormalization
method, for $\omega_c=0.3$ and for $e=1$.}
\end{figure}

Figure~\ref{fig:fomp} show a typical critical function
$\omega\mapsto F_c^+(\omega)$ obtained by the renormalization method 
for $\omega_c=0.3$ and for $e=1$. This figure is analogous to
Fig.~\ref{fig:fomn}.

From Fig.~\ref{fig:fomp}, we deduce that for $F>F_c=0.021$, no
invariant tori are left in this region of phase space. For these values of the
parameters $e$ and $\omega_c$, the condition of existence of invariant tori
$\omega\geq \sqrt{2}\omega_c\Vert V\Vert$ obtained from the integrable
case~(\ref{eqn:sint}), is $\omega\geq 0.67$. We notice that there is a large
chaotic zone for $\omega\in [0.67,0.8]$ (the set of $\omega$ where 
$F_c^+(\omega)$ is very small). The diffusion of trajectories throughout the $J$ 
large region is prevented by the small set of invariant tori with frequency 
$\omega\in [0.85,0.90]$ for intermediate values of the amplitude of the 
microwaves $F_c\in [0.015,0.02]$. The frequency of the last invariant torus to 
survive in this region of phase space is approximately equal to 0.87. The 
frequency of the last invariant torus in the positive twist region fluctuates in 
a very erratic way as one varies $\omega_c$. However we
observed that this frequency is always between 0.85 and 0.90 (close to the 1:1
resonance).

\begin{figure}
\centerline{\includegraphics*[scale=0.4]{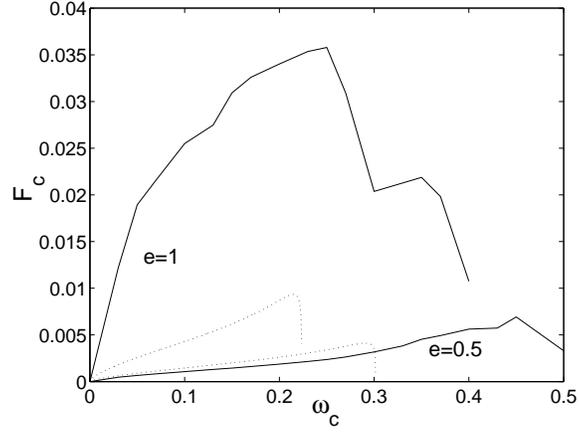}}
\caption{\label{fig:renchirp} Chaos thresholds $F_1^+(\omega_c)$ between 
resonances 1:1 and 2:1 in the positive twist region, obtained
 by the 2/3-rule criterion (dashed curves) and by the renormalization
method (continuous curves), as a function of $\omega_c$.  The upper
curves are obtained for $e=1$ and the lower ones are for $e=0.5$.}
\end{figure}

The changes induced by the magnetic field are
stronger in the positive twist region. We compute the critical
thresholds between 1:1 and 2:1 resonances in the positive twist
region. Figure~\ref{fig:renchirp} represents the results obtained by
renormalization and by the 2/3-rule criterion. In this region, the
critical thresholds determined by 2/3-rule criterion are well below
the ones given by the renormalization for significant values of the
field.

\begin{figure}
\centerline{\includegraphics*[scale=0.4]{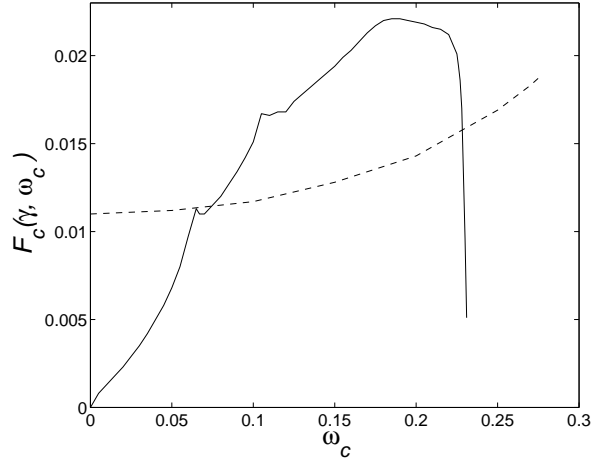}}
 \caption{\label{fig:Tgamma} Critical threshold $F_c(\gamma,\omega_c)$ for the 
break-up of the invariant tori with frequency $\gamma=(\sqrt{5}-1)/2$ for $e=1$ 
in the positive twist region (continuous curve) and in the negative twist
region (dashed curve).} 
\end{figure}

Figure~\ref{fig:Tgamma} represents the critical threshold for the break-up
of invariant tori with frequency $\gamma=(\sqrt{5}-1)/2$.
In the negative twist region, the golden mean torus is slightly stabilized by 
the magnetic field. In the positive
twist region, the influence of the field is very strong~: first, the torus
is created then stabilized to high values of the field, then it
disappears. This figure shows that in contrast to the negative twist
region, the influence of the magnetic field in the positive twist
region is very strong. The curve for the positive twist torus  appears to have
non-smooth variations conversely to the negative twist torus like for instance
for $\omega_c\approx 0.065$. We notice that for
the golden mean tori, there is no break-up by collision  since the one located in 
the positive twist region is broken before the expected collision.

For small values of the field $\omega_c$, we expand the threshold
given by Chirikov criterion ~(\ref{eqn:critchir}). Since the
resonances located at $J_n^+$ do not exist in the absence of the field
$\omega_c$, we expect $F_c^+(m,e,\omega_c)$ to vanish at $
\omega_c=0$. The corrections
to $F_c^+$ due to the magnetic field are obtained using
the following expansion of $
J_n^+=2^{1/3}n^{-1/3}\Vert V \Vert^{-2/3} \omega_c^{-2/3} +O(1)$
Thus we have~:
$ F_m^+(e,\omega_c)=\alpha_m(e)\omega_c^{2/3}[1+O(\omega^{2/3})]$,
where $\alpha_m$ are some constants depending on $e$. This means that
the increase of stabilization is very sharp (with infinite slope) for
low values of $\omega_c$. \\

\begin{figure}
\centerline{\includegraphics*[scale=0.4]{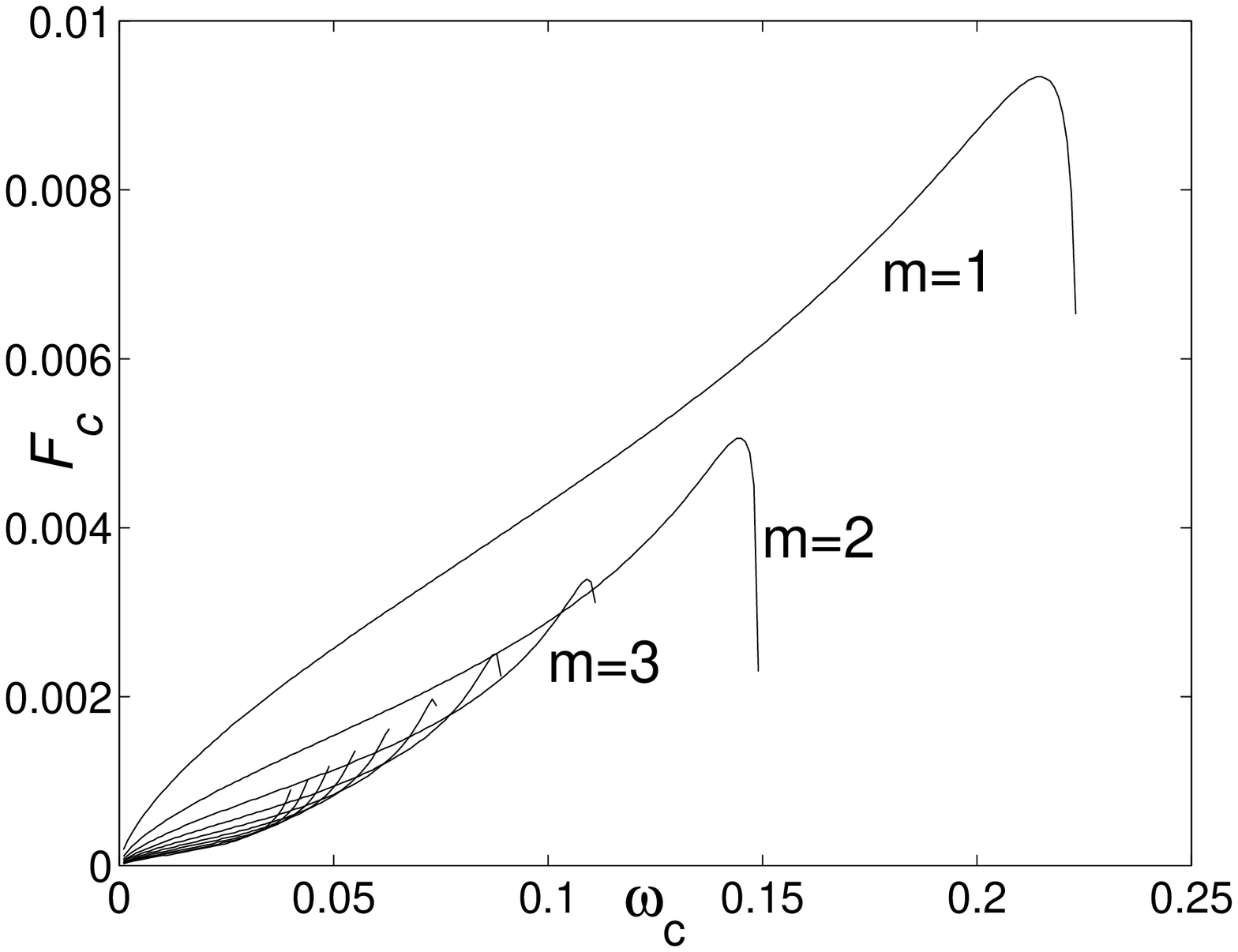}}
\caption{\label{fig:Epschirp} Resonance overlap values of 
$F_c^+(\protect\omega_c)$ between resonances $m$:1 and $m$+1:1 for $m=1,\ldots 
,10$ in the positive twist region for $e=1$.}
\end{figure}

Furthermore, using Eq.~(\ref{eqn:critchir}), we compute the 2/3-rule resonance
overlap value of $m$:1 and $m$+1:1 resonances for increasing $m=1,\ldots,10$.
For a fixed eccentricity $e=1$ , Fig.~\ref{fig:Epschirp} shows the
resonance overlap values as a function of the field $\omega_c$ in the positive
twist region.

What emerges from this figure is that this region of phase space
is strongly stabilized by the magnetic field, and that the curve $m=1$
dominates. However, this feature may vary with the parameter $e$ as it has been
observed in Ref.~\cite{chan02a} and in Sec.~\ref{sec:neg} for the negative
twist region. 

These curves are similar to Fig.~\ref{fig:fce}. As a result, increasing the magnetic
field makes intermediate eccentricity orbits as easy to ionize (diffusion
through a part of phase space where $J$ is large) as the ones with eccentricity
close to one. This situation is different from the situation without magnetic 
field. 

The overall effect of the magnetic field in the positive twist region is 
basically the same as the one in the negative twist region, i.e., it breaks up 
the primary resonances and fills up the remaining region by very stable 
invariant tori, preventing the diffusion of trajectories throughout phase space.

\section{Conclusion}

We have analyzed the classical phase space of the hydrogen atom in crossed
magnetic and microwave fields in the high frequency regime. Useful information 
about the dynamics is provided by the analysis of an integrable part of the 
Hamiltonian. Accurate information about chaos threshold is obtained by using the 
renormalization map and the 2/3-rule Chirikov criterion. The global effect of 
the magnetic field is to stabilize the dynamics and consequently reducing the 
diffusion of trajectories and the stochastic ionization process.


\newpage

\begin{figure}
\centerline{\includegraphics*[scale=0.4]{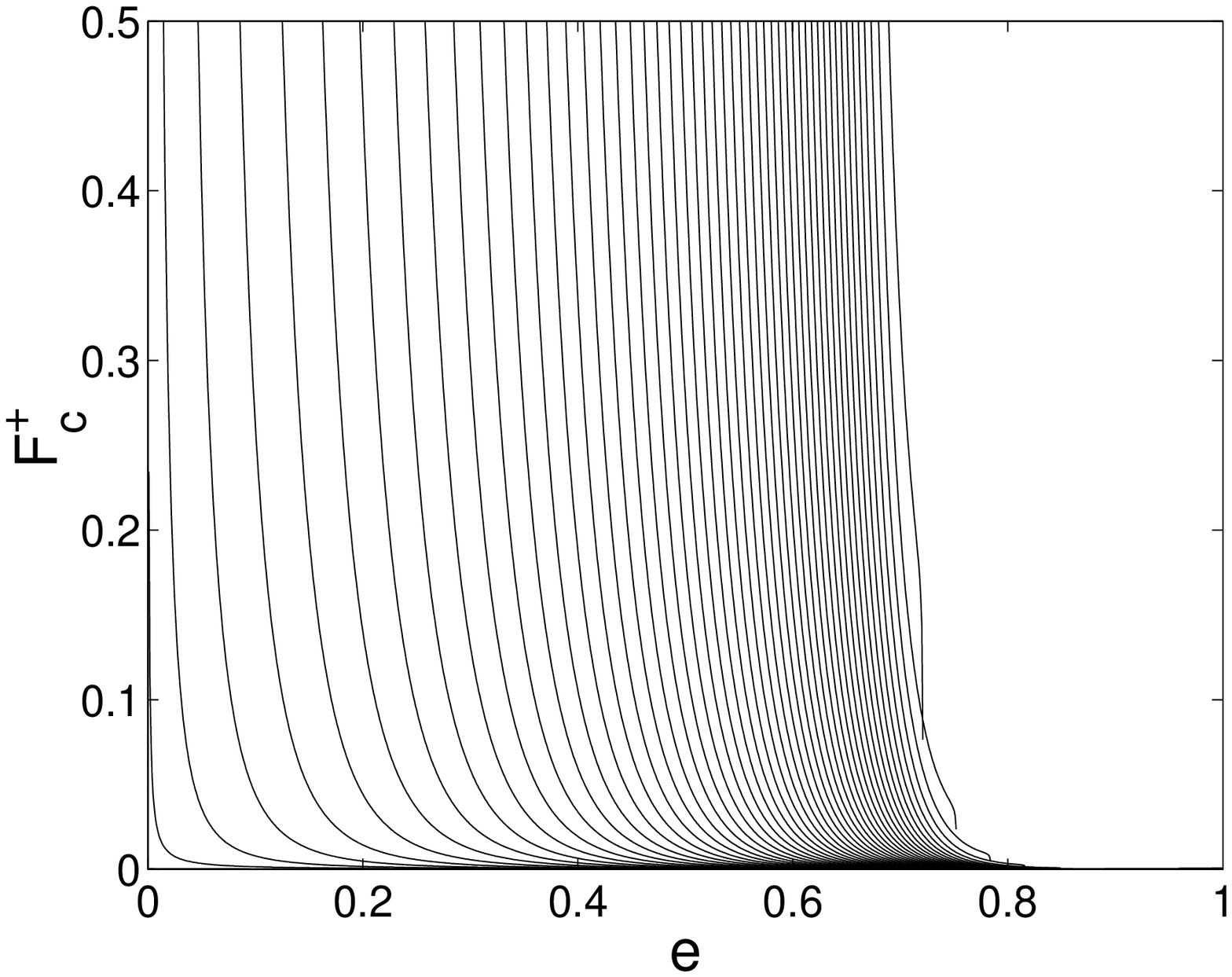}
\includegraphics*[scale=0.4]{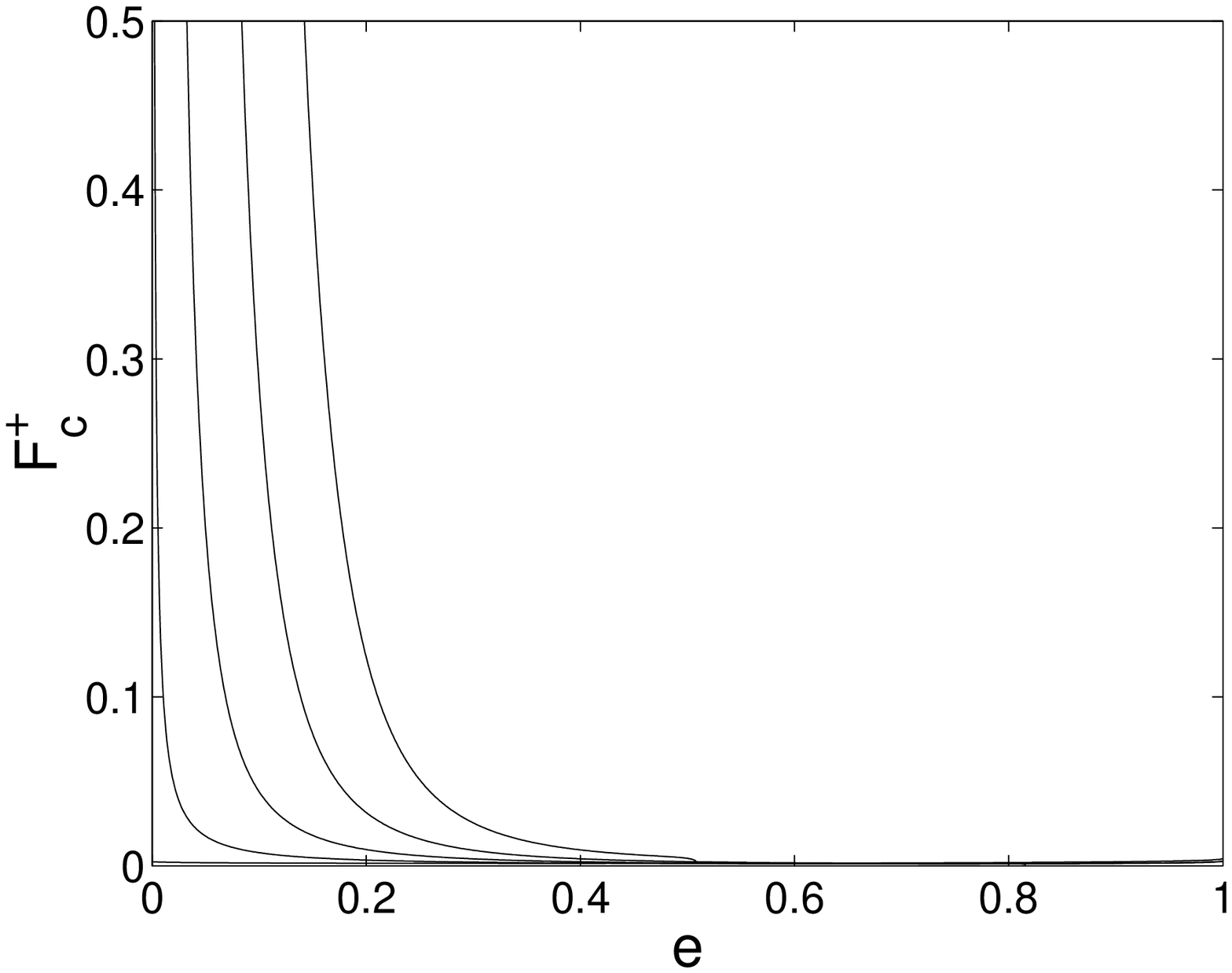}}
\caption{\label{fig:fcep} Resonance overlap values of $F_m^+(e)$ between
resonances $m$:1 and $m$+1:1 for $m=1,\ldots$ in the negative
twist region for $(a)$ $\omega_c=0.01$ and $(b)$ $\omega_c=0.1$.}
\end{figure}

\end{document}